\documentclass[fleqn,usenatbib]{mnras}
\usepackage{amsmath}
\usepackage{graphicx}
\usepackage{amssymb,txfonts}
\usepackage{xspace}

\newcommand{\msun}{{\rm M}_{\sun}}
\newcommand{\rsun}{{\rm R}_{\sun}}
\newcommand{\lsun}{{\rm L}_{\sun}}

\newcommand{\source}{{GX 339--4}\xspace}

\newbox\grsign \setbox\grsign=\hbox{$>$} \newdimen\grdimen \grdimen=\ht\grsign
\newbox\simpropbox
\setbox\simpropbox=\hbox{\raise.5ex\hbox{$\propto$}\llap
 {\lower.5ex\hbox{$\sim$}}}\ht2=\grdimen\dp2=0pt
\def\simprop{\mathrel{\copy\simpropbox}}

\title[The X-ray binary GX 339--4]{The X-ray binary GX 339--4/V821 Ara: the distance, inclination, evolutionary status and mass transfer}
\author[A. A. Zdziarski et al.]{Andrzej A. Zdziarski,\thanks{E-mail:
aaz@camk.edu.pl, jz@camk.edu.pl, mikolaj@camk.edu.pl} Janusz Zi\'o{\l}kowski$^\star$ and Joanna Miko{\l}ajewska$^\star$\\
Nicolaus Copernicus Astronomical Center, Polish Academy of Sciences,
Bartycka 18, PL-00-716 Warszawa, Poland}

\begin{document}

%\date{Accepted 2017 May 2. Received 2017 May 2; in original form 2017 February 25}

\pagerange{\pageref{firstpage}--\pageref{lastpage}} \pubyear{2019}

\maketitle

\label{firstpage}

\begin{abstract}
We consider constraints on the distance, inclination and component masses in the X-ray binary GX 339--4 resulting from published works, and then construct detailed evolutionary models for the donor. From both considerations, and assuming the black-hole nature for the compact object (i.e., its mass $>3\rm{M}_{\odot}$), the possible donor mass is $\approx$0.5--$1.4\rm{M}_{\odot}$, the inclination is $\approx{40}^\circ$--$60^\circ$ and the distance is $\approx$8--12\,kpc. The corresponding mass of the compact object is $\approx$4--$11\rm{M}_{\odot}$. We then confirm a previous estimate that the theoretical conservative mass transfer rate in GX 339--4 is $\lesssim{10^{-9}}\msun$\,yr$^{-1}$. This is $\gtrsim$10 times lower than the average mass accretion rate estimated from the long-term X-ray light curve. We show that this discrepancy can be solved in two ways. One solution invokes irradiation of the donor by X-rays from accretion, which can temporarily enhance the mass transfer rate. We found that absorption of a $\sim$1 per cent of the irradiating luminosity results in the transfer rate equal to the accretion rate. The time scale at which the transfer rate will vary is estimated to be $\sim$10\,yr, which appears consistent with the observations. The other solution invokes non-conservative mass transfer. This requires that $\approx$70 per cent of the transferred mass escapes as a strong outflow and carries away the specific angular momentum comparable to that of the donor.
\end{abstract}
\begin{keywords}
binaries: general -- stars: evolution -- stars: individual: V821
Ara-- stars: low-mass -- X-rays: binaries -- X-rays: individual: GX
339--4.
\end{keywords}

\section{Introduction}
\label{intro}

\source was discovered by {\it OSO-7\/} satellite in 1972 as a new X-ray transient \citep{markert73}. The source was soon noticed for its high variability. Already \citet{markert73} found that it changed its luminosity by a factor $\approx$60 in one year. Its optical counterpart has got the variable-star designation V821 Ara. Since the discovery, the source, classified as a soft X-ray transient, has shown more frequent (several per decade) outbursts (observed from radio to soft $\gamma$-rays) than any other transient low-mas X-ray binary (LMXB; see \citealt*{coriat12} for a review). Apart from that, GX 339--4 behaves as a typical black-hole (BH) LMXB. In particular, it displays all X-ray states observed in other BH LMXBs \citep{mendez97}.

The current best determination of its binary parameters is that by \citet{heida17} (hereafter H17). The orbital period is $P= 1.7587 \pm 0.0005$ d (hereafter the uncertainties quoted from H17 are $1\sigma$), confirming and refining the original determination by \citet{hynes03}. The mass function is $1.91 \pm 0.08\msun$ and the ratio the donor mas, $M_2$, to the accretor mass, $M_1$, is $q = 0.18 \pm 0.05$. H17 then considered the lack of eclipses, which gives an upper limit on the inclination. This limit and the mass ratio imply the minimum compact object mass of $M_1\approx 2.9\pm 0.3\msun$. The distance to GX 339--4 has been estimated by \citet{zdziarski04} (hereafter Z04) as $7\,{\rm kpc}\lesssim D\lesssim 9\,{\rm kpc}$, while H17, using a similar method, found $D\gtrsim 5$ kpc, with a preference for a larger distance. 

\citet*{munoz08} (hereafter MD08) calculated theoretical rates of the mass outflow from the donor using equations of \citet{king93} (which were, in turn, based on the formulae of \citealt*{webbink83}). They obtained the maximum allowed rate of $-\dot M_2\approx 7.8\times 10^{-10}\msun$ yr$^{-1}$, and concluded that this value agrees with the transient nature of the source. However, MD08 have not estimated the observed average accretion rate on the BH. That was done by Z04 and \citet{coriat12} based on the X-ray observations covering jointly a few decades. Assuming the accretion efficiency of 0.1 and $D=8$ kpc, the average mass accretion rate is rather high, $\dot M_1\approx (1.3$--$1.4) \times 10^{-8} \msun\, {\rm yr}^{-1}$. As noted by \citet{basak16}, these values are by a factor $\sim$15 higher than the maximum value of MD08. 

In this paper, we first reconsider the constraints on the system masses, distance and inclinations using the results of H17 and \citet{buxton12}. We then develop evolutionary models for the donor, and determine their implied binary parameters. The models also give us rates of the conservative mass transfer from the donor, which confirm both the estimates of MD08 and the presence of the large discrepancy between the theoretical transfer rate and the observational accretion rate. We then study possible resolutions of this major discrepancy between the theory and observations. We investigate whether the discrepancy can be resolved by the effect of irradiation of the donor by X-rays from accretion. As an alternative solution, we consider a non-conservative mass transfer in the binary.

\section{Model-independent constraints}
\label{constraints}

We consider here some model-independent constraints resulting from the observational data. H17 measured the rotation velocity along the line of sight as $v_{\rm rot}\sin i \approx 64\pm 8$ km s$^{-1}$. This implies the donor radius of 
\begin{equation}
R_2=\frac{P\, (v_{\rm rot} \!\sin i)}{2\upi \sin i}\approx \frac{(2.22\pm 0.28)\rsun}{\sin i}\gtrsim \frac{1.95\rsun}{\sin i}.
\label{radius}
\end{equation}
The measured range of values of $v_{\rm rot}\sin i$ also yields\footnote{We note that the relationship \citep{gies86} $v_{\rm rot}/K_2=(1+q)R_{\rm L}/A$ is commonly used with the approximation to $R_{\rm L}/A$ of \citet{paczynski67}, valid for small $q$. For an arbitrary value of $q$, the approximation $R_{\rm L}/A\approx 0.49 q^{2/3}/[0.6 q^{2/3}+\ln(1+q^{1/3})]$ \citep{eggleton83} can be used.} the mass ratio of $q\approx 0.18^{+0.06}_{-0.05}$. The lack of eclipses with the assumptions of the Roche-lobe filling and corotation of the donor with the binary imply\footnote{Here, we use an approximation of the donor being a sphere with the Roche lobe radius, implying $i_{\rm max}=\arccos(R_{\rm L}/A)$. H17 gave $i_{\rm max}=78\degr$, which probably takes into account the flattening of the Roche lobe.} $i\leq 77\degr$, corresponding to the lowest allowed value of $q$ (and thus the minimum measured $v_{\rm rot}\sin i$). The Roche lobe radii for $M_2\lesssim 0.6 M_1$ are well approximated \citep{eggleton83} by a formula by \citet{paczynski67}, $R_{\rm L}/A\approx (2/3^{4/3})(1+1/q)^{-1/3}$ (where $A$ is the binary separation), which, when combined with the Kepler law, yield
\begin{equation}
R_L\approx (2 G M_2)^{1/3} (P/9\upi)^{2/3}.
\label{p67}
\end{equation}
We assume $R_2=R_{\rm L}$ hereafter. Note that the above approximation gives a relationship between $R_2$ and $M_2$ independent of $M_1$. Using the minimum $v_{\rm rot}\sin i \approx 56$ km s$^{1}$, the above relationships imply
\begin{equation}
R_2\geq 2.0\rsun,\quad M_2\geq 0.35\msun.
\label{radius2}
\end{equation}
They are important model-independent constraints, which we give here for the first time. 

Obviously, at a given donor mass, $M_2$ (and thus $R_2$), possible values of $i$ and $q$ are not independent of each other, but instead positively correlated, 
\begin{equation}
\sin i=\frac{P K_2}{3^{4/3}\upi R_2}q^{1/3}(1+q)^{2/3}.
\label{i_vs_q}
\end{equation}
Consequently, the minimum values of $R_2$ and $M_2$, corresponding to the eclipse limiting value of $i=77\degr$, also correspond to a narrow range of the values of $q$ close to its minimum, with the uncertainty corresponding to the uncertainties of the measured radial velocity, $K_2$, and the period, $P$, which yield $q\approx 0.13$--0.14. Then, the corresponding mass of the compact object is $M_1\approx 2.6$--$2.7\msun$. Given the similarity of \source to other BH binaries, these values of $M_1$ appear unlikely. Furthermore, the shape of the track of \source on the X-ray hardness-count rate diagram implies $i\lesssim 60\degr$ \citep{munoz13}. 

We can also use constraints given by flux densities. Assuming a blackbody emission, we have the observed flux density of
\begin{equation}
F_\nu=\upi B_\nu(T_{\rm eff}) \frac{R_2^2}{D^2},
\label{bb}
\end{equation}
where $B_\nu(T_{\rm eff})$ is the blackbody intensity at the effective temperature $T_{\rm eff}$. Its value for the donor in \source remains uncertain. H17 found three best-fitting templates, which were by the spectra of the K2 III star HD 175545 and the K1 IV star HD 165438, and by a synthetic atmosphere model with $T_{\rm eff}=3938$\,K, $\log_{10} g=2.5$, and the solar metallicity. Among them, HD 175545 gave the lowest $\chi^2$ of their fits while the 3938\,K model gave the strongest cross-correlation between the observed and template spectra. The latter would appear to suggest a relatively low temperature of the donor in \source. However, we note that the surface gravity of that template, a parameter strongly affecting the appearance of stellar spectra, is an order of magnitude lower than that in our case,
\begin{equation}
g\equiv \frac{GM_2}{R_2^2}=\frac{(9\upi)^2 R_2}{2 P^2}\approx 3012(R_2/2.5 \rsun)\,{\rm cm\,s}^{-2}.
\label{gravity}
\end{equation}
On the other hand, $T_{\rm eff}$ and $\log_{10} g$ of the stars providing the best-fitting templates are $\approx\! 4500\pm 50$\,K, $\approx$3.0 in HD 175545 \citep*{koleva12, mcdonald17} and $\approx\! 4900\pm 50$\,K, $\approx$3.4 in HD 165438 \citep*{prugniel11,jones11, reffert15, luck17}, respectively. Those values of $\log_{10} g$ are significantly closer to the range of $\approx$3.4--3.6 of the possible models (Section \ref{model}). Therefore, we hereafter adopt the range of 4400--5000\,K as that possible in the donor. This implies the intrinsic stellar flux density in the middle of the $H$ band ($1.62\,\mu$m; least affected by extinction) of $F_{\nu\star H} \approx 0.15(R_2/2.5\rsun)^2 (D/10\,{\rm kpc})^{-2} (T_{\rm eff}/4500\,{\rm K})^{2.4}$\,mJy, where the exponent of 2.4 has been fitted to the dependence in the 4000--5000 K range.

\begin{figure}
\centerline{\includegraphics[width=7.5cm]{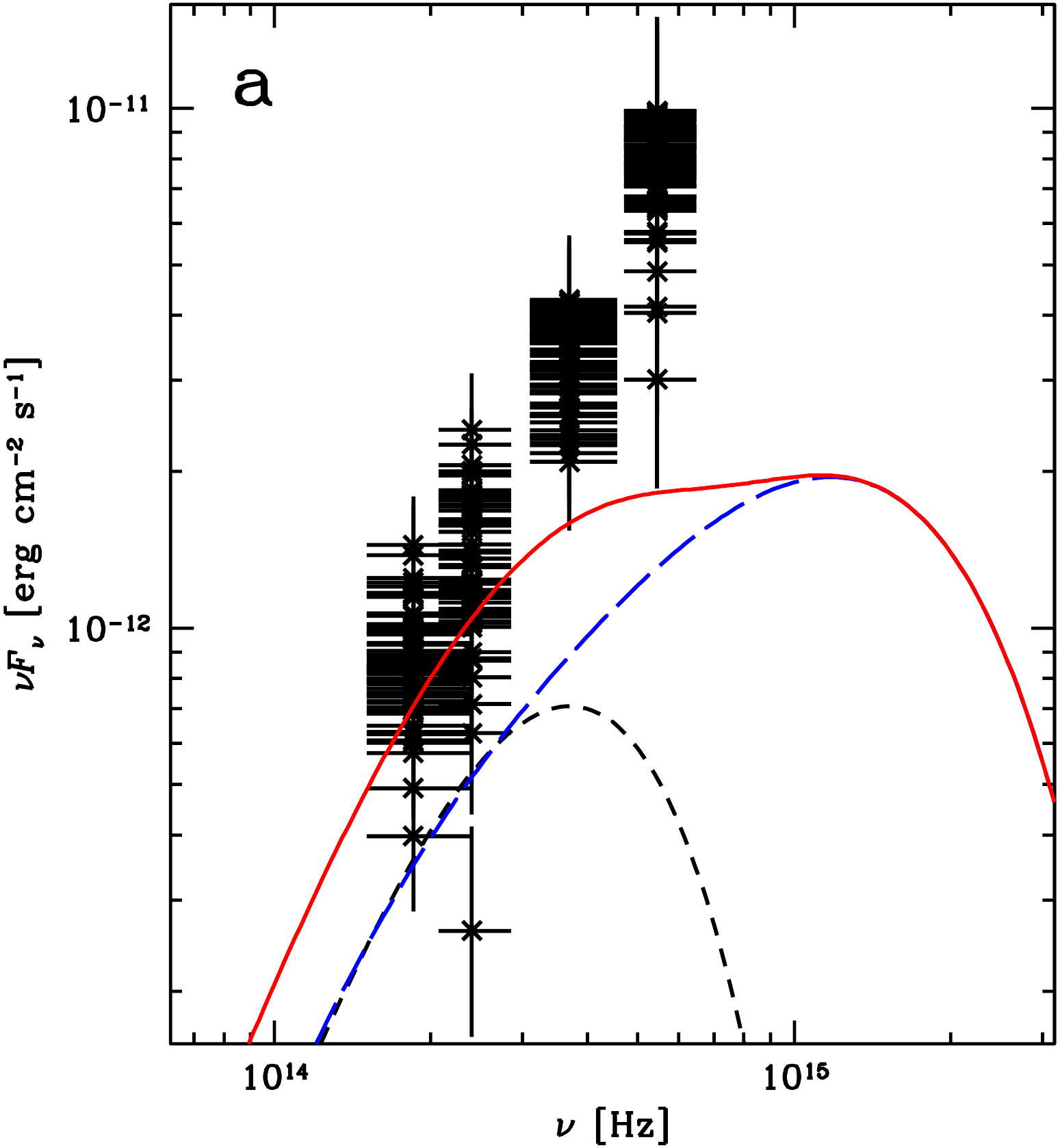}}
\centerline{\includegraphics[width=7.5cm]{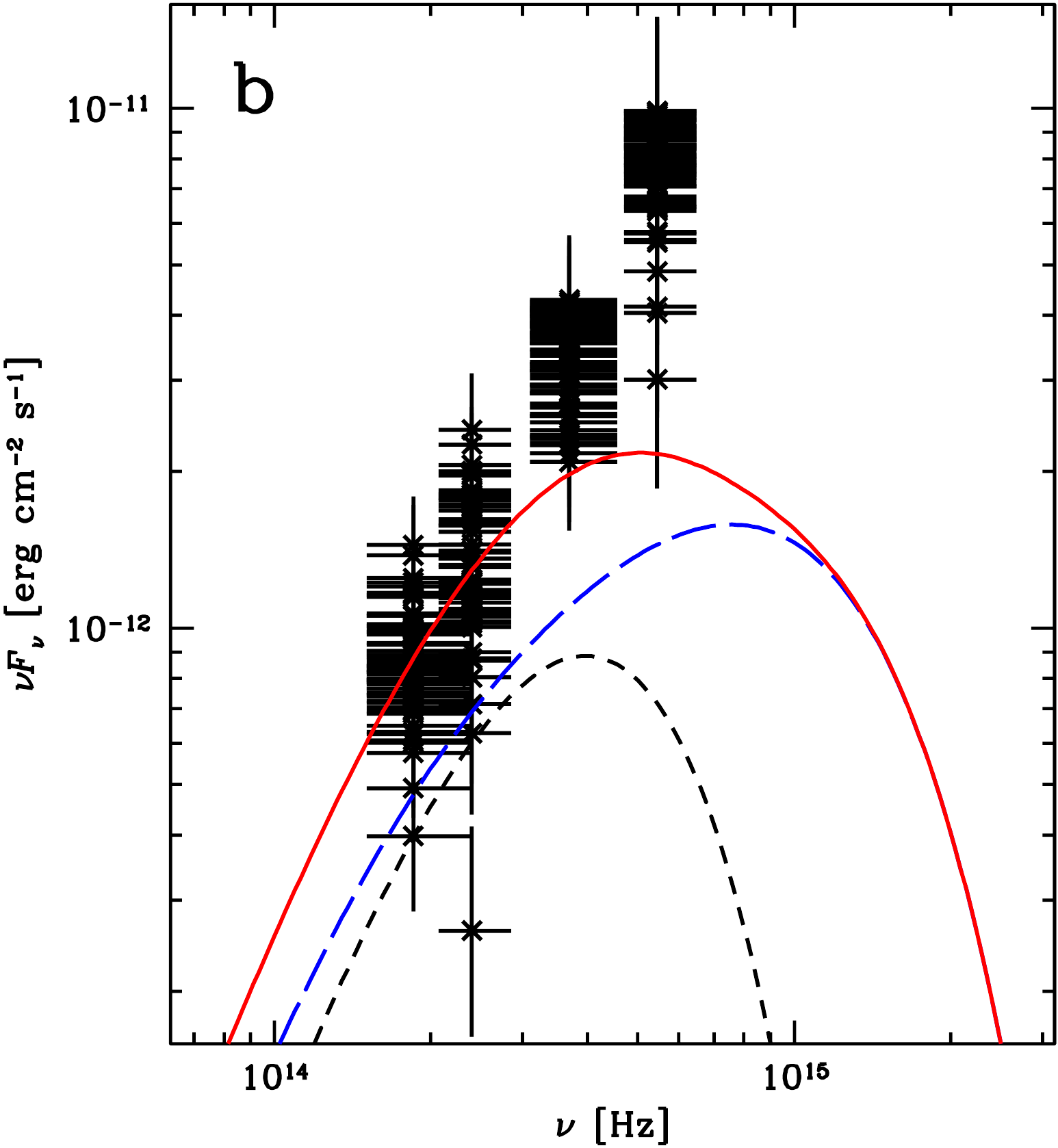}}
\caption{IR--O spectra of \source in quiescence. The black error bars give the lowest measurements of \citet{buxton12} selected by the criterion of $I >18$ and $V >19.5$. The uncertainties include the $1\sigma$ magnitude uncertainties and the systematic uncertainties in $E(B-V)$ and an assumed 10 per cent uncertainty in the conversion from the magnitude to flux. The black short-dashed curves show stellar blackbody spectra, and the blue long-dashed curves give disc spectra. The red solid curves give the sum. The disc is modelled as blackbody from $R_{\rm out}=0.8 R_{\rm L,accretor}$ at $T_{\rm out}=2000$\,K to $R_{\rm in}=2\times 10^4 R_{\rm g}$ with $T\propto R^{-3/4}$, $q=0.13$. The model parameters are: the stellar mass and spectrum of (a) model B (see Table \ref{models}), $D=8$ kpc, $i=60\degr$ and (b) model E, $D=11$ kpc, $i=40\degr$. See Section \ref{constraints} for details.
}
 \label{spectra}
 \end{figure}

We then study the lowest IR fluxes measured in \source using the results of \citet{buxton12}, who performed intensive monitoring of \source in the $V,\, J,\, I$ and $H$ bands during 2002--2010. Their analysis was concentrated on the outburst states of the source, and while they give the results of all of their observations, no discussion of the quiescence was given. Their lowest measured $H$ magnitudes of $>$17 are at the sensitivity limit of their 1.3 m telescope. Therefore, we correlated the $H$ magnitudes against those of $J,\,I$ and $V$. We have found, in particular, that the $H$ magnitudes approach on average a constant value (with significant scatter) at $I>18$. Thus, we select the quiescent state based on this criterion, and in addition, we also impose a criterion of $V>19.5$. The reddening towards the source is approximately $E(B-V)\approx 1.2\pm 0.1$ \citep{zdziarski98}. The fluxes corrected for the extinction are shown\footnote{We use the Cousin's system and define the middle wavelengths of the $V,\, J,\, I$ and $H$ bands at 0.55, 0.812, 1.25, $1.62\mu$m, their fractional band widths as 0.16 0.19 0.16 0.23, and the fluxes corresponding to zero magnitudes as 3750, 2680, 1615, 1090 Jy, respectively. We use the extinction, $A_\lambda$, as equal to 3.1, 1.76, 0.87, 0.59, respectively, times the reddening, $E(B-V)$.} by black error bars in Fig.\ \ref{spectra}. From that, we infer that the lowest reliable detections in the $H$ band correspond to an average of $F_\nu\approx 0.4$ \,mJy, or $\nu F_\nu\approx 7.5\times 10^{-13}$ erg s$^{-1}$ cm$^{-2}$. We then assume, similarly as found by H17 for their observations, that a half of that flux, i.e., $F_{\nu\star H}\approx 0.2$ mJy, comes from the star. This is also consistent (in spite of possible substantial systematic errors) with the three weakest $H$ magnitudes recorded by \citet{buxton12} if we assume they were dominated by the stellar contribution. They are $17.26\pm 0.16$, $17.31\pm 0.17$ and $17.47\pm 0.19$, which give the extinction-corrected fluxes of $F_\nu \approx 0.26^{+0.09}_{-0.07},\, 0.25^{+0.09}_{-0.07},\, 0.22^{+0.08}_{-0.07}$\,mJy, respectively. The above considerations imply a distance dependent on $F_{\nu\star H}$, $R_2$ and $T_{\rm eff}$,
\begin{equation}
D\approx 6.9\,{\rm kpc}\left(\frac{F_{\nu\star H}}{0.2\,{\rm mJy}} \right)^{-1/2} \frac{R_2}{2\rsun} \left(\frac{T_{\rm eff}}{4500 \, {\rm K}}\right)^{1.2},
\label{distance}
\end{equation}
The value of $R_2$ above can be substituted by that of equation (\ref{radius}), giving the constraint dependent on the binary inclination.

Fig.\ \ref{spectra} also shows some example spectra from the star and the accretion disc. The shown stellar spectra use the parameters of of models B (at 8 kpc) and E (at 11 kpc) of Section \ref{model}, and their fluxes at the middle frequency of the $H$ band are 0.19 and 0.21 mJy, respectively. The shown disc spectra use the model of \citet{ss73}, with the blackbody spectrum at $T\propto R^{-3/4}$ integrated over the disc surface. The parameters are $M_1$ corresponding to $q=0.13$, $i=40$ and $60\degr$, respectively, the outer edge of the disc of at 0.8 of the Roche-lobe radius of the BH, the temperature at that edge of 2000 K, and the inner disc truncation radius \citep*{dubus01, bernardini16} of $2\times 10^4 R_{\rm g}$, where $R_{\rm g}\equiv G M_1/c^2$. 

On the other hand, H17 gave their observed $H$-band flux as $\approx$0.11 mJy. Since their spectroscopic data have no photometric calibration, they estimated its uncertainty to be by a factor of two. Based on their stellar templates, they estimated the contribution of the accretion disc during their observations as being of $\approx$50 per cent. Taking the above estimates into account, the extinction-corrected stellar flux density in the $H$ band is $F_{\nu\star H}\approx 0.10^{+0.10}_{-0.05}$\,mJy. Thus, their maximum measurement estimate of $F_{\nu\star H}$ agrees with our estimate of $\approx$0.2\,mJy, and we use this value hereafter.

We note that the constraint on $D$ of equation (\ref{distance}) is very similar to that obtained by Z04 based on the maximum allowed Gunn $r$-band ($0.67\mu$m) magnitude of the donor of \citet*{shahbaz01} (as corrected in Z04). While that Z04 in their estimates used the previous, apparently overestimated, measurement of the mass function by \citet{hynes03}, those authors also gave a value of $q\lesssim 0.08$, lower than that of H17, which then lead to the estimate of of Z04 of $R_2\approx 2.3\rsun/\sin i$, which is almost identical to the current determination of equation (\ref{radius}). Thus, the constraints on $D$ of Z04 remain approximately valid, with an upward correction for the donor temperature being higher than 4000\,K assumed by them. On the other hand, H17 obtained $D\gtrsim 5$\,kpc, but that estimate was based on the minimum-mass model of MD08, with $M_2=0.166\msun$, which is less than a half of the minimum allowed mass (equation \ref{radius}). When rescaling to the actual minimum mass, their results are fully consistent with ours.

Equation (\ref{distance}) implies the lower limit on the distance of $D\gtrsim 7$\, kpc. This limit corresponds to both the maximum inclination of $77\degr$ set by the lack of eclipses and the minimum of $v_{\rm rot}\sin i$, which then corresponds to the compact object mass of (2.6--$2.7)\msun$. Both $i=77\degr$ and this range of mass appear unlikely (as we point out above). This indicates that the actual distance is higher than 7 kpc. On the other hand, the upper limit on the distance depends on a minimum allowed inclination through $R_2(i)$ of equation (\ref{radius}), which is model-dependent, and which we discuss in Section \ref{model} below.

\begin{table*}
\caption{The results of the evolutionary calculations for the donor of \source and the corresponding constraints imposed by the binary solution. Approximately, $\sin i\propto M_1^{-1/3}$ at a given $M_2$, see equation (\ref{i_vs_q}). The distance scales as $D\propto (F_{\nu\star H}/0.2\,{\rm mJy})^{-1/2}$, see equation (\ref{distance}). The Eddington flux from accretion onto the compact object scales with as $F_{\rm Edd}\propto (F_{\nu\star H}/0.2\,{\rm mJy})^{-1}$  (its stated uncertainty corresponds to the dependence of $F_{\rm Edd}\propto M_1$). The models denoted by a single letter have the solar metallicity of $Z=0.014$, while models B2, C2, D2 have twice that, $Z=0.028$. The surface gravity, $g$, is in units of cm s$^{-2}$.
}
\begin{center}
\label{models}
\begin{tabular}{cccccccccccc}
\hline Model &
$\displaystyle{\frac{M_2}{\msun}}$ &
$\displaystyle{\frac{R_2}{\rsun}}$ &
$\displaystyle{\frac{L_2}{\lsun}}$ &
$T_{\rm eff}[{\rm K}]$ &
$\log_{10} g$&
$i[\degr]$ &
$\displaystyle{\frac{M_1}{\msun}}$ &
$\frac{\displaystyle{D\,[{\rm kpc}]}}{
\left(\frac{F_{\nu\star H}}{0.2\,{\rm mJy}} \right)^{-1/2}}$ &
$\frac{\displaystyle{F_{\rm Edd}[10^{-8}{\rm erg\,cm}^{-2}{\rm s}^{-1}]}}{\left(\frac{F_{\nu\star H}}{0.2\,{\rm mJy}} \right)^{-1}}$ &
$\displaystyle{\frac{M_{\rm c}}{\msun}}$ & 
$\displaystyle{\frac{-\dot M_2}{10^{-10} \msun\,{\rm yr}^{-1}}}$\\
\hline
A &0.35 &2.00 &1.47 &4520 & 3.38 &77--77 & 2.6--2.7 &7.0 &6.4--6.8 & 0.142& 1.16 \\
B &0.50 &2.25 &1.89 &4530 & 3.43 &60--76 & 2.8--3.9 &7.9 &5.5--7.6 & 0.138& 2.55 \\
B2&0.50 &2.25 &1.49 &4268 & 3.43 &60--76 & 2.8--3.9 &7.3 &6.3--8.8 & 0.149& 1.5  \\
C &0.70 &2.52 &2.55 &4616 & 3.48 &51--75 & 3.1--5.5 &9.0 &4.6--8.2 & 0.148& 4.95 \\
C2&0.70 &2.52 &2.08 &4389 & 3.48 &51--75 & 3.1--5.5 &8.5 &5.2--9.2 & 0.145& 3.9  \\
D &1.00 &2.83 &3.68 &4770 & 3.53 &43--62 & 4.1--7.8 &10.5 &4.5--8.6 & 0.148& 9.75\\
D2&1.00 &2.83 &3.04 &4546 & 3.53 &43--62 & 4.1--7.8 &10.0 &5.0--9.5  & 0.152& 7.0\\
E &1.20 &3.01 &4.46 &4855 & 3.56 &40--56 & 4.9--9.4 &11.4 &4.6--8.8 & 0.156& 11.4\\
F &1.40 &3.17 &5.40 &4962 & 3.58 &38--52 & 5.8--11.0&12.3 &4.6--8.8 & 0.155& 17.0\\
\hline
\end{tabular}
\end{center}
\end{table*}

\section{The evolutionary model}
\label{model}

\begin{figure}
\centerline{\includegraphics[width=7.cm]{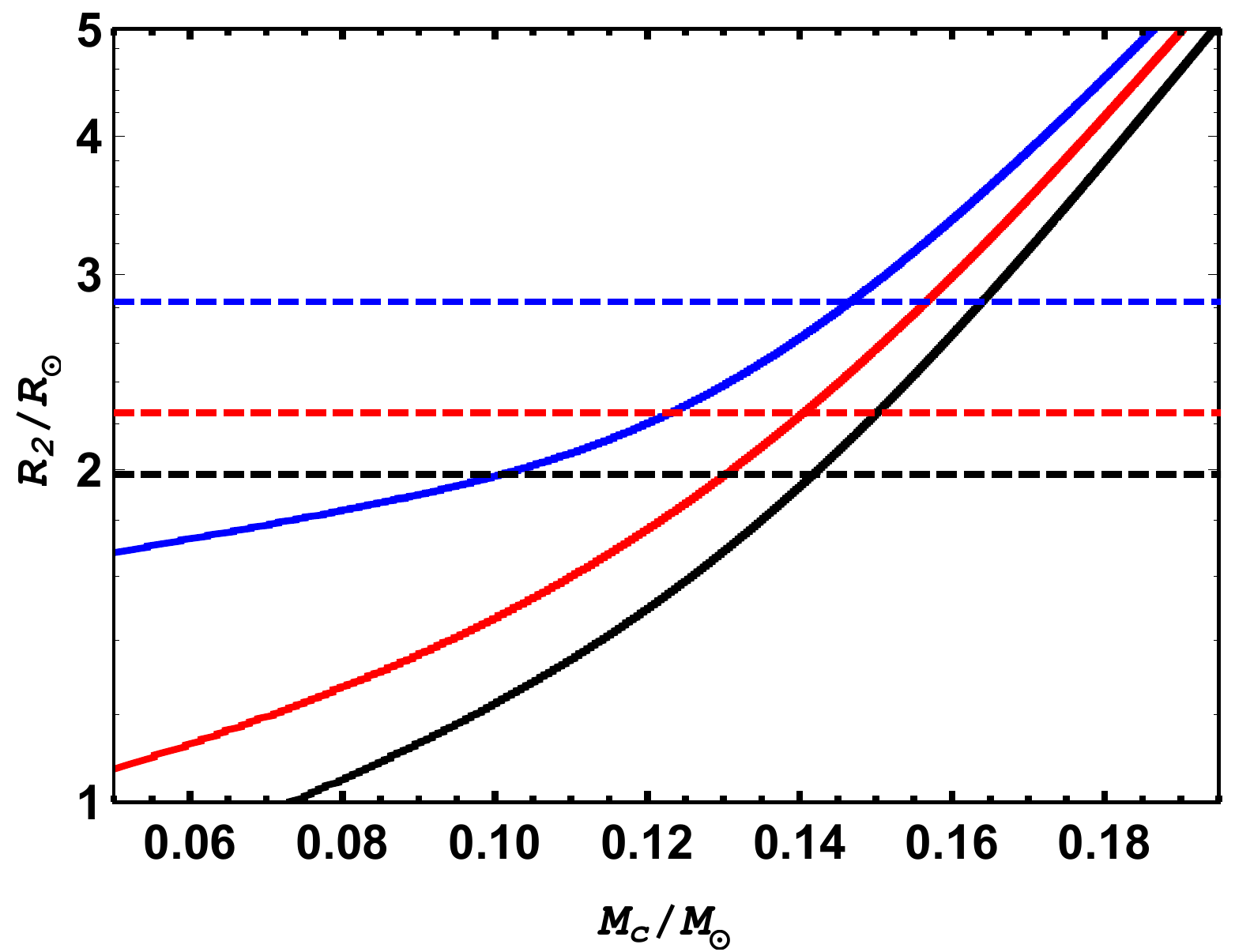}}
\caption{Evolution of partially stripped giants for total masses of $M_2=0.35$, 0.5 and $1.0 \msun$ shown (from bottom to top) by the black, red and blue solid curves, respectively. The evolution proceeds from left to right at the constant $M_2$. The horizontal dashed lines with the corresponding colours (from bottom to top) show the radii of the Roche lobe around the donor for the above values of $M_2$. The crossings of the corresponding evolutionary tracks and horizontal lines determine the positions of our donor models. We see that these crossing correspond to a narrow range of $M_{\rm c}\approx (0.14$--$0.15)\msun$. 
}
 \label{evolution}
 \end{figure}

We use the Warsaw stellar evolution code of \citet{paczynski69,paczynski70}, also developed by M. Koz{\l}owski and R. Sienkiewicz. Its main current features and updates (e.g., the used opacities, nuclear reaction rates, equation of state) are described in \citet{pamyatnykh98} and \citet{ziolkowski05}. In \citet{z16}, the code was calibrated to reproduce the Sun at the solar age. This resulted in the H mass fraction of $X=0.74$, the metallicity of $Z=0.014$, and the mixing length parameter of 1.55. Here, we use it to follow evolution of stripped giants, following the approach used to calculate models of IGR J17451--3022 \citep{z16}, GRS 1915+105/V1487 Aql \citep{zz17} and V404 Cyg/GS 2023+338 \citep{zz18}.

In order to reproduce a system close to \source, we followed the evolution of a $2 \msun$ main-sequence star, which was maintained at a constant mass until hydrogen was nearly exhausted in its centre. Then, the mass removal from the surface started and it was continued until the donor star reached a specified value of the mass. The He core still did not form by that time. We found that the assumed rate of the mass removal did not affect the final results. The further evolution of the remnant was followed at a constant $M_2$. The H-burning shell moved outwards, increasing the mass of the formed He core, $M_{\rm c}$, and decreasing that of the H-rich envelope. We calculate not only the radius and the luminosity of each model but also its entire internal structure, which is essential for calculating the reaction of the star to mass transfer.

The results of our evolutionary calculations for \source are given in Table \ref{models}. For a given $M_2$, $R_2$ follows from the Roche lobe formula, and $M_1$ and the range of inclinations follow from the measured $K_2$ and $v_{\rm rot}$. The other quantities, in particular $L_2$ and $M_{\rm c}$ follow from the evolutionary calculations. Fig.\ \ref{evolution} shows the evolutionary tracks for the stellar masses of $M_2=0.35, 0.5$ and $1.0\msun$ of stripped giants in the core mass, $M_{\rm c}$, vs.\ $R_2$ plane. In addition, we also consider models with $M_2=0.7$, 1.2, and $1.4\msun$, which are not shown in Fig.\ \ref{evolution}. The stars evolve at the constant mass and the driving mechanism is the progress of the H-burning shell moving outwards. The radii of the partially stripped giants generally increase with $M_{\rm c}$, except for the shrinking when the masses of their envelopes become very low. Note also that the He cores we consider here have sufficiently small masses for the dependence of the giant parameters not only on the core mass but also on the total mass to be substantial. The crossings of the evolutionary tracks with the corresponding radii of the donor Roche lobe give the possible models. We find that there are no solutions for $M_2\lesssim 0.2 \msun$ since the corresponding giants never attain sufficiently large radii during their evolution (not shown in Fig.\ \ref{evolution}). However, the limiting mass of $0.2\msun$ is below the observational minimum $M_2\approx 0.35\msun$, given by equation (\ref{radius2}). The models with $M_2=0.35$, 0.5, 0.7, 1.0, 1.2 and $1.4\msun$, are labelled A, B, C, D, E and F, respectively. Since the metallicity of the donor in \source is not known, we have also considered the metallicity of twice solar, $Z=0.028$, for models B, C and D, labelled B2, C2, D2, respectively. The parameters of all of the models are given in Table \ref{models}. Table \ref{models} also give the allowed ranges\footnote{The lower limit on the mass of the compact object is given by $q_{\rm max}$ being the minimum of the maximum value of $q$ from the rotational broadening and the solution of the eclipse-limit condition of $(R_{\rm L}/A)(1+q)(K_2-\Delta K_2)(P-\Delta P)/(2\upi R_2)=\sqrt{1-(R_{\rm L}/A)^2}$, where $R_{\rm L}/A$ is a function of $q$ (approximated by the formulae of either \citealt{paczynski67} or \citealt{eggleton83}). The corresponding upper limit on $i$ is given by $i_{\rm max}=\arccos(R_{\rm L}/A)$ at $q_{\rm max}$.} of $i$ and $M_1$ (based on equation \ref{i_vs_q}), and the scaling for $D$ (based on equations \ref{bb} and \ref{distance}). Furthermore, Table \ref{models} gives the bolometric Eddington flux from accretion onto the compact object, assuming isotropy and the solar H fraction,
\begin{equation}
F_{\rm Edd}\equiv \frac{G c m_{\rm p} M_1}{\sigma_{\rm T} D^2} \frac{2}{1+X},
\label{Edd}
\end{equation}
where $m_{\rm p}$ is the proton mass and $\sigma_{\rm T}$ is the Thomson cross section.

%\vbox to 2cm{\vfill}
%a line to avoid pdflatex fatal error

Model A has $M_1\approx 2.6$--$2.7\msun$, which corresponds to either a very heavy neutron star or a very light BH. Such low BH mass has never been observed, and theoretical calculations of dense nuclear matter indicate that the maximum mass of a neutron star is $< 2.4 \msun$ \citep{lp10}. Currently, the largest neutron star mass measured at high accuracy is $2.01\pm 0.04\msun$ \citep{antoniodasis12}. Recently, objects with higher masses, but determined to a lower accuracy, have been found. The values of $2.27^{+0.17}_{-0.15}\msun$ and $2.17^{+0.11}_{-0.10}\msun$ have been obtained from measurements of the radial velocities (including detailed corrections for irradiation by the donor; \citealt*{linares18}) and of the Shapiro delay \citep{cromartie19}, respectively. On the other hand, the mass of $2.40\pm 0.12\msun$ found by \citet{vankerkwijk11} appears less reliable, as discussed, e.g., in \citet{ozel16}. Thus, neutron star masses of 2.6--$2.7\msun$ appear unlikely.

In fact, the properties of \source are very similar to a number of confirmed BH binaries, which suggests the presence of a BH with a mass higher than the above range. Also, as noted above, the shape of the tracks of the X-ray hardness-count rate diagram of \source is similar to transient BH binaries with known low inclinations, $\lesssim 60\degr$ or so, but clearly different from those with the inclination measured to be $\gtrsim 70\degr$ \citep{munoz13}. That dependence on the inclination is well explained by the anisotropy of the flux from the geometrically-thin accretion disc in this system. This further disfavours model A. 

Based on our adopted temperature criterion of $T_{\rm eff}$ in the 4400--5000\,K range (see Section \ref{constraints}), model B2 is disfavoured. Model B appears acceptable at the lowest value of $q=0.13$ allowed by the measurement of the rotational broadening, which corresponds to $i=60\degr$ (and thus satisfying the criterion of \citealt{munoz13}) and $M_1=3.9\msun$. While this BH mass is low, it cannot be ruled out. Similar considerations apply to models C and C2, which have lower inclinations and higher BH masses; at $q=0.13$, they have $i=51\degr$ and $M_1=5.5\msun$, which appear fully acceptable. Models D--F fully satisfy the inclination condition of $i\gtrsim 60\degr$. Their allowed BH mass ranges overlap with the BH-transient observed range of $7.8\pm 1.2\msun$ of \citet{ozel10}. On the other hand, \citet{parker16}, based on X-ray spectral fits, found $M_1\approx (8$--$12)\msun$ and $D\approx 8$--10 kpc (including both the statistical and systematic errors), which mass range favours $M_2\gtrsim 1\msun$, while their distance range favours models with $0.5\lesssim M_2/\msun\lesssim 1$. Model F with $M_2=1.4\msun$ still satisfies (approximately) our temperature criterion, and we thus consider $M_2\approx 1.4\msun$ as the highest possible mass in the binary. Nevertheless, given that H17 found that the $\approx$4500\,K stellar template provides the overall best fit, a lower donor mass is preferred.

MD08 estimated a similar maximum allowed mass of the donor, $1.1\msun$, based on the constraint of $M_{\rm c}\geq 0.17 M_2$ assumed by them to be required for the donor to be a giant. This limit is of \citet{sc42}, and it corresponds to the maximum possible mass of a stellar core that is still isothermal and in hydrostatic equilibrium; for higher masses, the core contracts and the star moves towards the red giant branch. However, the ratio of $M_{\rm c}/M_2$ for that limit given in literature is in the $\approx$0.10--0.17 range \citep*{beech88,eggleton98, ball12}, and our own evolutionary calculations yield it at $M_{\rm c}/M_2\approx 0.1$. Our models E and F with $M_2=1.2$ and $1.4\msun$ have $M_{\rm c}/M_2\approx 0.13$ and 0.11, respectively, i.e., they satisfy this limit. On the other hand, quasi-stationary solutions obviously exist below this limit, with the star being then evolved (as a subgiant) but with the core still isothermal. This happens, e.g., for our solution with $M_2=2\msun$, which has $M_{\rm c}/M_2\approx 0.05$. However, we rejected that solution due to its very high $T_{\rm eff}$, and therefore it is not shown in Table \ref{models}.

\section{The mass transfer rate}
\label{transfer}

 \begin{figure}
\centerline{\includegraphics[width=6.5cm]{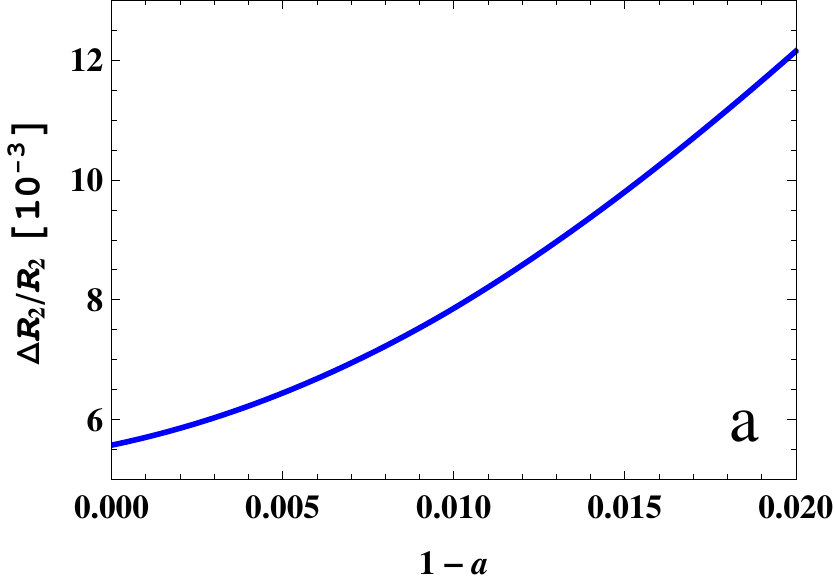}}
\centerline{\includegraphics[width=6.5cm]{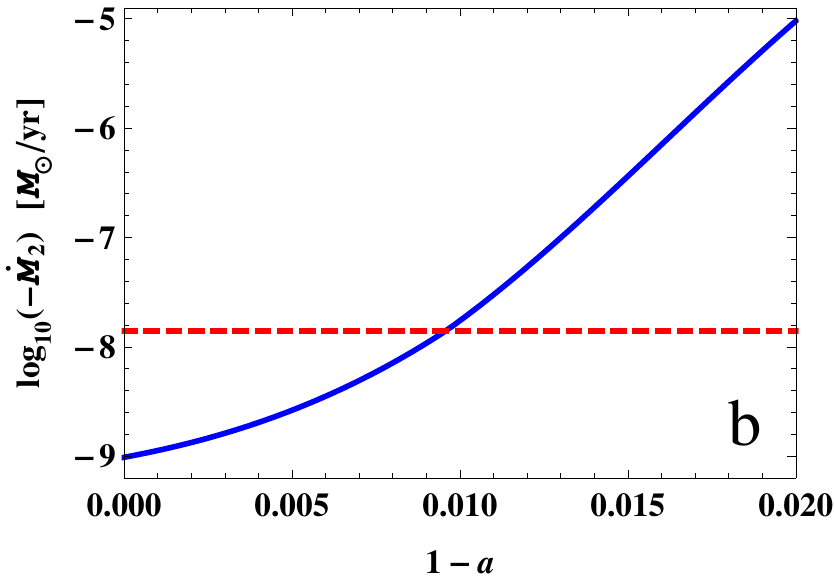}}
 \caption{The dependence of (a) the radius excess of the donor and (b) its mass outflow rate on the X-ray albedo, $a$. We show the dependencies as functions of $1-a$. The red horizontal line in (b) shows the observed average $\dot M_1$, which should equal $-\dot M_2$ in the conservative model. See Section \ref{irrad} for details.}
 \label{deltaR}
 \end{figure}

As we stated in Section \ref{intro}, the average accretion rate on the compact object estimated from the average X-ray luminosity is quite high, $\dot M_1\approx (1.3$--$1.4) \times 10^{-8} \msun\,{\rm yr}^{-1}$ (Z04; \citealt{coriat12,basak16}). On the other hand, the evolutionary model of the donor proposed by MD08 predicted the rate of the outflow from the donor of only $\approx 8\times 10^{-10}\msun$ yr$^{-1}$. Confirming that, the models described in Section \ref{models} also predict small outflow rates, thus much smaller than those inferred from the observations. We assume the BH mass of $M_1=M_2/q_{\rm min}$, where $q_{\rm min}=0.13$ is the minimum mass ratio of H17, and, in Sections \ref{irrad}--\ref{oscill}, the conservation of the total mass and the total orbital angular momentum. Then, we calculate numerically at which rate of the mass outflow from the star the changes of the stellar radius will follow the changes of the Roche lobe around it. The resulting rates are given in Table \ref{models}, where we see a fast increase of $-\dot M_2$ with increasing $M_2$. The highest value of $-\dot M_2\approx 1.7\times 10^{-9}\msun$ yr$^{-1}$ was obtained at $1.4\msun$ (model F). Hereafter in this Section, we use $M_2=1\msun$ (model D), yielding only slightly lower $-\dot M_2\approx 1.0\times 10^{-9}\msun$ yr$^{-1}$.

\subsection{Irradiation of the donor}
\label{irrad}

It was noted by many authors that strong irradiation of the donor by X-rays from accretion onto the compact object is likely to cause the donor outer envelope to expand. This expansion will induce a faster mass outflow. Estimates of this effect were done, among others, by \citet{webbink83}, \citet{podsiadlowski91}, \citet*{frank92}, \citet{HKLR93}, \citet*{VEF94}, \citet{KFKR96,KFKR97}, \citet*{RZK00} and \citet{ritter08}. Results obtained by different authors were substantially different. We have decided to obtain a simple estimate of the expansion effect following \citet{webbink83}, as given by their equation 16. We have estimated the effective temperature of the heated hemisphere of the donor, $T_{\rm heated}$,
\begin{equation}
T_{\rm heated}^4 = T_{\rm intr}^4+ \frac{(1-a) L_{\rm irrad}}{2 \upi
\sigma R_2^2}, \label{Te}
\end{equation}
where $T_{\rm intr}$ is the effective temperature of the donor in the absence of irradiation, $L_{\rm irrad}$ is the irradiating flux, and $a$ is the X-ray albedo of the stellar surface. We have then constructed a model of the irradiated envelope starting from the heated photosphere. We have done such calculations for different values of the albedo using the average bolometric accretion luminosity of $\langle L\rangle \approx 8\times 10^{37}$ erg s$^{-1}$ (assuming $D=8$ kpc). The irradiating flux is then calculated from the geometry of the system. We find that the energy falling on the heated hemisphere makes up $\approx$1.2 per cent of the bolometric accretion flux, which yields the average $L_{\rm irrad} \approx 260\lsun$. On the other hand, the intrinsic luminosity of the donor is only $\approx 2.8$--$3.7\lsun$ at $1\msun$, and less for lower masses, see Table \ref{models}.

We then used the formula of \citet{jed69} as given in \citet{ps72} to relate the radius excess of the donor, $\Delta R_2$, i.e., the difference between the irradiated donor radius and the Roche lobe radius, to the rate of the mass outflow from the donor. It gives the relation $-\dot M_2 \propto (\Delta R_2/R_2)^{n+3/2}$, where $n$ is the local polytropic index at the Roche lobe surface, but the form of the proportionality coefficient is complicated. The values of $n$ for our models are in the range of $\sim$3--6. We found that in order to obtain the outflow rate predicted by the internal structure of the adopted unirradiated model (D; $9.75\times10^{-10} \msun$ yr$^{-1}$), we need $\Delta R_2/R_2 \approx 5.6 \times 10^{-3}$. We note that in order to calculate it, we have had to iterate because the proportionality coefficient depends both on $n$ and the entropy of the matter at the Roche lobe surface. The mass in the excess layer above the Roche lobe surface and the density at the Roche lobe surface were found to be $\approx 3.0 \times 10^{-7} \msun$ and $\rho = 3.1\times 10^{-6}$ g cm$^{-3}$, respectively. The thermal scale height at the Roche lobe surface was found to be $\approx 3.6 \times 10^9$ cm and was comparable to the thickness of the layer ($\approx 1.1\times 10^9$ cm).

We then assumed that the mass contained in the excess layer above the Roche lobe surface does not change when irradiated but the layer only expands. The results of the calculations of the expanded structure of the irradiated envelope are shown in Fig.\ \ref{deltaR}(a). We have found that the thickness of the excess layer does, as expected, increase with the increasing irradiation (decreasing albedo), but this increase is relatively moderate. We have found that $\Delta R_2/R_2$ changes from $\approx 5.6 \times 10^{-3}$ for $a = 1$ to $\approx 1.2 \times 10^{-2}$ for $a = 0.98$. The corresponding values of the rate of the mass outflow from the donor are shown in Fig.\ \ref{deltaR}(b). We see that they depend very strongly on the albedo; increasing $1-a$ from 0 to 0.02 increases the rate of the outflow by four orders of magnitude. Fig.\ \ref{deltaR}(b) shows that in order to obtain $-\dot M_2 = \dot M_1$ we need $1 - a = 0.0096$, i.e., the donor has to absorb $\approx$1 per cent of the irradiating flux.

\subsection{The oscillatory behaviour of the rate of the mass outflow}
\label{oscill}

The irradiation effect discussed above can increase the outflow rate only on a relatively short timescales. The outflow on long timescales is still determined by the internal structure of the donor, i.e., by the rate of H burning on the surface of the He core. Therefore, irradiation of the donor might lead to oscillatory behaviour of the rate of the mass outflow, i.e., cycles of the mass transfer. This effect was analysed by \citet{KFKR96,KFKR97} and discussed at some length by \citet{ritter08}. Their conclusion was that while the mass transfer cycles may develop in many classes of LMXBs, the time scales of these cycles are determined by the time scales of the donor outer convective envelope, which means that they are rather long. 

On the other hand, we notice, based on Z04 and \citet{coriat12}, that the X-ray fluxes of GX 339--4 observed over more than twenty years do show an increase of the average flux on the time scale of a decade. Z04 notice that this might be an effect of the feedback from the donor irradiation. We consider here the possibility that this is related to the time scale of emptying an element of the outflowing matter near the $L_1$ point. We follow the approach of \citet{Savonije83} and \citet*{ZWG07}. The perpendicular cross section of a flow through $L_1$ is given by, e.g., equation 10 of \citet{ZWG07}, 
\begin{equation}
{\cal A}\approx \frac{P^2 G M_2 \Delta R_2}{2 \upi R_2^2}. 
\label{area}
\end{equation}
For our binary parameters at $M_2=1\msun$, ${\cal A}\approx 1.26\times 10^{13} (\Delta R_2/1\,{\rm cm})$ cm$^2$. For the unirradiated model, $\Delta R_2 \approx 5.6 \times 10^{-3} R_2 \approx 1.1\times 10^9$ cm, and $\rho$ is given above. The volume of our patch is therefore $V\approx A\Delta R_2 \approx 1.5\times 10^{31}$ cm${^3}$. The mass of our element is therefore $\Delta M \approx V\rho \approx 4.7 \times 10^{25}$ g. Given $-\dot M_2$ predicted for unirradiated model is $\approx 6.1\times 10^{16}$ g s$^{-1}$, the time scale for emptying the characteristic $L_1$ volume is $\Delta t\equiv \Delta M/\dot M_2 \approx 24$ yr.

Let us consider now the irradiated case, where $\Delta R_2/R_2\approx 7.7 \times 10^{-3}$, and thus $\Delta R_2 \approx 1.5\times 10^9$ cm. This gives $V \approx 2.9\times 10^{31}$ cm${^3}$. The density at the Roche lobe surface for our model is $2.5\times10^{-6}$ g cm$^{-3}$, implying $\Delta M \approx 7.2\times 10^{25}$ g. The outflow rate, $-\dot M_2$, calculated for irradiated model is assumed equal to the observed average accretion rate of $\approx 1.4\times10^{-8} \msun$ yr$^{-1}$, which implies $\Delta t\approx 2.6$ yr.

Even if the convection within the stellar envelope is effective in mixing the surface layers with its deeper parts \citep{KFKR96,KFKR97,ritter08}, the outflowing region around $L_1$ may respond relatively fast to irradiation, and the above time scales are likely to be relevant for the variability of the outflow rate from the donor. They are, to the order of magnitude, in agreement with the observed time scale of long-term variability of the accretion rate. 

\subsection{The non-conservative mass transfer}
\label{nonconservative}

\begin{figure}
\centerline{\includegraphics[width=7.cm]{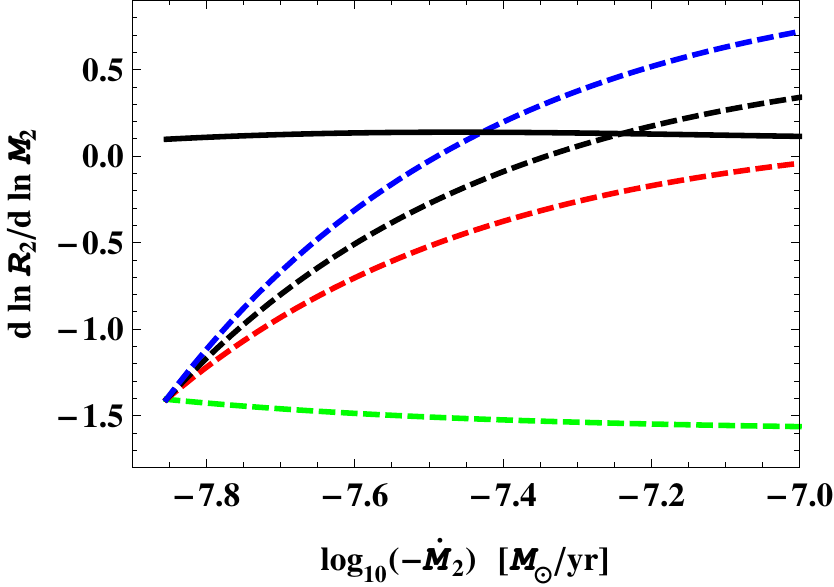}}
\caption{The rates of the evolutionary change of the radius of the donor and that of its Roche lobe, ${\rm d}\ln R_2/{\rm d}\ln M_2$, as functions of the donor mass loss rate, $-\dot M_2$, for non-conservative mass transfer. The black solid curve gives the results of our evolutionary model including the effect of mass outflow from the outer layers of the star. The dashed curves show the rates of the Roche lobe change for values of $\alpha$ increasing from bottom to top; $\alpha=0$ (green), $\alpha=1$ (red), $\alpha=1.25$ (black), $\alpha=1.5$ (blue). The intersections of the solid curve and the dashed curves give possible self-consistent models of the source, at $\beta=-\dot M_1/\dot M_2$ (with $\dot M_1= 1.4\times10^{-8} \msun$ yr$^{-1}$), and the origin of the dashed curves is at $\beta=1$. See Section \ref{nonconservative} for details.}
\label{drdm}
 \end{figure}

Here, we discuss an alternative solution allowing us to increase the outflow rate from the donor. This solution relies on non-conservative mass transfer. As customary in this case \citep{verbunt93}, we use the fraction of the mass lost by the donor that is accreted, $\beta$, and the specific angular momentum of the mass leaving the system divided by the specific angular momentum of the donor (measured from the centre of mass), $\alpha$. We look for possible solutions by comparing the rates of the evolutionary change of the radius of the donor and of its Roche lobe as functions of the donor mass loss rate, $-\dot M_2$. The evolutionary changes of the radius of the donor were computed by imposing the outflow at different rates for $M_2$ and $R_2$ of our model D. The results are shown with the red solid curve in Fig.\ \ref{drdm}. The rate of the change of the Roche lobe radius vs.\ $M_2$ was calculated by \citet{rappaport82}. When expressed as a function of $\alpha$ and $\beta$, it is given by equation 6 of \citet{zz18},
\begin{equation}
\frac{{\rm d}\ln R_2}{{\rm d}\ln M_2} = -\frac{5}{3} +2\beta
\frac{M_2}{M_1} + \frac{2}{3}(1-\beta) \frac{M_2}{M_1+M_2} +2 (1-\beta)
\alpha \frac{M_1}{M_1+M_2}. \label{RL}
\end{equation}
In the present case, the model mass outflow rate from the donor is larger than the accretion rate, opposite to the conservative case without irradiation.

The results for different values of $\alpha$ are shown with the green dotted curves in Fig.\ \ref{drdm}. The curves originate at the conservative solution, with $\beta=1$ and $-\dot M_2=\dot M_1= 1.4\times 10^{-8}\msun$ yr$^{-1}$ and $M_2=1\msun$, $q=0.13$. The red solid curve shows ${\rm d}\ln R_2/{\rm d}\ln M_2$ for our evolutionary model including the mass transfer, and we see that this rate at $-\dot M_2=\dot M_1$ disagrees with that for the Roche lobe response. However, this changes for higher values of $-\dot M_2$, and self-consistent solutions for a given $\alpha$ correspond to the intersections of the two curves. At the intersections, $\beta=-\dot M_1/\dot M_2$. The shown intersections correspond to ($\alpha=1.5,\,\beta \approx 0.38$) and ($\alpha=1.25,\,\beta \approx 0.30$).

We need then to consider what values of $\alpha$ are expected during the mass transfer. While values of $\alpha$ of several are found for outflows through $L_2$ (during contact phases of a binary), see, e.g., \citet{flannery77}, its values for the outflows from the vicinity of $L_1$ and from the viscously formed accretion disc are much lower, as discussed in detail in \citet{zz18}. The maximum value of $\alpha_{\rm max}$ is achieved at the disc outer radius, which we assume to be 0.9 of the accretor Roche lobe radius. Then, $\alpha_{\rm max}$ is solely a function of the mass ratio, and can be calculated using equations 7--11 of \citet{zz18} and the approximation to the Roche lobe radius of \citet{eggleton83}. We find $\alpha_{\rm max}$ is a weak function of $q$, with $\alpha_{\rm max}\approx 1.19$, 1.23, 1.29 for $q=0.13$, 0.18, 0.23, respectively. This is comparable to $\alpha=1.25$ considered in Fig.\ \ref{drdm}. Thus, the non-conservative transfer appears possible for $\alpha\approx 1.25$ and $\beta\approx 0.3$. 

However, we note that unlike the case of V404 Cyg showing very strong winds, which motivated the study of the non-conservative mass transfer of \citet{zz18}, no evidence for disc outflows has so far been found in GX 339--4. In particular, high-ionization Fe K absorption lines in the soft state, which trace disc winds, have not been detected \citep{ponti12}.

\section{Period change}
\label{Pdot}

The mass transfer rate in a binary is associated with a period change. We use the standard formula, which, when expressed in terms of $\alpha$ and $\beta$, is given by equation 15 in \citet{zz18}. In the conservative case with irradiation causing a temporary increase of the mass transfer rate, $-\dot M_2=\dot M_1\approx 1.4\times 10^{-8}\msun\,{\rm yr}^{-1}$ for $M_2=1\msun$, $M_1\approx 7.7\msun$, we have the characteristic time scale for period increase of $P/\dot P\approx 2.7\times 10^7$\,yr. In the non-conservative case with $-\dot M_2\approx 1.4\times 10^{-8}\beta^{-1}\msun\,{\rm yr}^{-1}$, $\alpha=1.25$, $\beta=0.3$ (as found in Section \ref{nonconservative}), $P/\dot P\approx 4.5\times 10^7$\,yr, i.e., it is longer by a factor of two from that of the conservative case. Given the uncertainties in the used parameters, these estimates should be considered as representative, and the possibility of distinguishing between the two scenario based on a future measurement of $\dot P$ appears uncertain.

In general, mass transfer alone from the less massive component to the more massive one leads to an increase of $\dot P$ for plasible values of the parameters; namely $\dot P> 0$ for $\alpha< 1+2q/3$ at $\beta=0$, and up to higher values of $\alpha$ at $\beta>0$. Orbital period changes have been measured in only three\footnote{$\dot P$ has also been measured in the short period ($P \approx 0.20$\,d) high-mass X-ray binary Cyg X-3, where $\dot P>0$ and $P/\dot P\approx 1.0\times 10^6$\,yr (e.g., \citealt{bhargava17}), which value appears to be in agreement with the rate of wind mass loss in that system. However, while it is likely that the compact object in that system is a BH, the presence of a neutron star remains possible \citep*{zmb13,koljonen17}.} accreting BH binaries, namely in the short period systems (with $P\approx 0.17$--0.43 d) XTE J1118+480, A0620--00 and Nova Muscae 1991 \citep{gonzalez17}. Contrary to the above expectation, $\dot P$ was found to be $<0$ in all three cases, with the best-fitted time scales of $-P/\dot P\approx 7.7\times 10^6$, $4.6\times 10^7$, $1.8\times 10^6$\,yr, respectively. It is possible that the negative $\dot P$ is due to the dominance of magnetic braking in those systems and their descendance from highly magnetized intermediate-mass binaries \citep*{justham06}, which, however, requires the current presence of rather high magnetic fields in the donors \citep{gonzalez17}. We note that while some variations of the value of $\dot P$ implied by the mass transfer and mass loss can be due to fluctuations of the value of $\dot M_2$ (see the last paragraph in Section \ref{discussion} below) and variable disc wind loss rates, a change of the sign of $\dot P$ cannot occur due to such fluctuations (see equation 15 in \citealt{zz18}).

Given the absence of a $\dot P$ measurement in \source, we have no indication whether or not magnetic braking operates in this source. Since its orbital period is much longer than that in the three systems discussed above, the donor may have evolved from a star with a low magnetic field \citep{justham06}, and magnetic braking may be not operative. However, we point out that if the actual $\dot P$ is substantially less than that of our estimates for \source, e.g., due to the presence of magnetic braking or another effect, the associate change of the rate of the orbit evolution would affect our estimates of $\dot M_2$. In particular, it would increase the estimates of $-\dot M_2$ in the conservative transfer case (Table \ref{models}). This effect thus represents a third hypothetical scenario, in addition to the irradiation and non-conservative mass transfer, that could reconcile the theoretical and observed values of $\dot M_2$ in \source.

\section{Discussion}
\label{discussion}

We have estimated the possible range of the distance to \source, but our estimates still bear a substantial uncertainty related to the unknown exact value of the donor flux. They could be improved with future studies of the photometry of the source in quiescence. Also, strong distance constraints could have been achieved if the spectroscopic observations of H17 were accompanied by sensitive photometry. We also note that in our estimates we assumed blackbody emission. Using stellar-atmosphere models instead would reduce somewhat the predicted fluxes at a given $T_{\rm eff}$, and would correspondingly slightly reduce the distance estimates.

Still, our results support relatively low values of the BH mass and large distances, as seen in Table \ref{models}. These values have then implications for the determination of the BH spin using the disc continuum spectral fitting method \citep{kolehmainen10}. Those authors used the disc atmosphere model of \citet{davis05}, which incorporates calculations of the disc vertical structure, radiative transfer and relativistic effects, and fitted simultaneously nine spectra of \source in the soft state. From their table 2, we find that our constraints imply low to moderate values of the dimensionless spin, $a_*\la 0.8$ (allowing for a misalignment of the inner disc with respect to the orbital plane).

Another uncertainty concerns the origin of the accretion spectrum in quiescence. As we see in Fig.\ \ref{spectra}, simple disc models together with the stellar emission can reproduce well the quiescent $H$ and $J$ measurements of \citet{buxton12}. However, the models fall below the observational points at $I$ and $V$. Furthermore, we note that the accretion disc in quiescence is not in a steady state, and thus the model of \citet{ss73} does not apply. However, reproducing the $I$ and $V$ data points would require the local temperature increasing with the decresing radius faster than $T\propto R^{-3/4}$ of the steady state. This is opposite to the theoretical expectations; e.g., in fig.\ 13 of \citet{dubus01} we see that the temperature profile in quiescence is flatter than $R^{-3/4}$. This would then give fluxes at the $I$ and $V$ bands much lower than those of our example model.

Thus, the question arises of the origin of most of the $I$ and $V$ fluxes of \source\ in quiescence. Apart from the optically-thick disc, the accretion flow consists of a hot inner flow downstream the inner disc truncation radius, e.g., \citet*{lasota96}. The hot electrons would then Compton-upscatter the disc emission, leading to a component extending from IR to X-rays. In addition, the flow can contain some non-thermal electrons, which then would lead to efficient synchrotron emission of the flow, e.g., \citet{pv14}. The combination of the two processes could account for the observed emission, but future studies are desirable.

In Section \ref{model}, we calculated the bolometric flux, $F_{\rm Edd}$, that would be observed from the accretion flow in \source isotropically emitting the Eddington luminosity, equation (\ref{Edd}). In principle, this could serve as a diagnostic allowing us to select preferable models. However, we find that due to a chance approximate proportionality of $D^2\simprop M_1$, the allowed ranges of $F_{\rm Edd}$ are very similar for all the considered models, $\approx (5$--$10)\times 10^{-8}$ erg cm$^{-2}$ s$^{-1}$. These values can be compared with observations. The highest measured bolometric flux, $F_{\rm bol}$, appears to be $\approx 5\times 10^{-8}$ erg cm$^{-2}$ s$^{-1}$ (Z04), observed in the soft state. This implies that \source is a source emitting up to around the Eddington limit. Notably, the highest $F_{\rm bol}$ seen in the hard state is only slightly lower, $\approx 3.5\times 10^{-8}$ erg cm$^{-2}$ s$^{-1}$ \citep{islam18}.  

We then studied the effect of irradiation of the donor on the mass transfer rate (Section \ref{irrad}), and found that it can increase the rate substantially. This then can resolve the discrepancy between the estimated transfer and accretion rates. Such an effect is also expected in other X-ray binaries. It may be most pronounced in \source given it is the most often outbursting source. Among other sources, \citet{shaw19} found that the mass transfer rate in the transient BH-candidate binary Swift J1753.5--0127 is variable during its very long outburst, which may be related to donor irradiation.

We note here that if donor irradiation is indeed present during quiescence, the so-called K correction to the mass function (due to the lines being emitted preferentially by the irradiated hemisphere) may become important. In their section 4.1, H17 discuss that correction and conclude it is unimportant given the lack of evidence for irradiation. However, if it does operate, the inferred mass function would increase, affecting the masses and inclination of the system.

Another possible solution of the rate discrepancy is via non-conservative mass transfer (Section \ref{nonconservative}). We note that the rate problem in \source is opposite to that in V404 Cyg, where the estimated mass transfer rate was much {\it higher\/} than the estimated accretion rate, and the main effect of the mass removal from the binary was a reduction of the accretion rate. In the present case, the non-conservative mass and angular momentum losses lead to an increase of the transfer rate. 

As a cautionary note, we point out that the change of the donor Roche lobe radius (which is driving $\dot M_2$) in 10 yr, which is a time scale on which the mass transfer rate can be measured, is only $\approx\! 10^5$\,cm, while the radius excess of the Roche lobe required for the theoretical rate in the unirradiated case (see Section \ref{irrad}) is $\approx\! 10^9$\,cm. Thus, relatively small fluctuations in the structure of the Roche lobe excess around $L_1$ may lead to strong fluctuations of $\dot M_2$. A quantitative analysis of this effect is, however, beyond the scope of our paper.

\section{Conclusions}

Our main results are as follows. 

We have determined current model-independent constraints on the distance, inclination and component masses in the X-ray binary GX 339--4, as following from H17 and \citet{buxton12}. The minimum radius and mass of the donor are $2.0\rsun$ and $0.35\msun$, respectively. However, those values correspond to unlikely values of the inclination of $77\degr$ and the compact object mass of (2.6--$2.7)\msun$. On the other hand, the masses of $M_2\gtrsim 0.5\msun$ yield BH masses of $\gtrsim 4\msun$ and allow the inclinations of $i\lesssim 60\degr$ (which agrees with the constraint on $i$ of \citealt{munoz13}). We have then constructed evolutionary models for the donor. Comparing those models with the observational data, we find the possible donor mass to be within the $\approx\! 0.5$--$1.4\msun$ range, the BH mass within $\approx\! 4$--$11\msun$, the inclination within $\approx\! 40\degr$--$60\degr$ and the distance within $\approx\! 8$--12\,kpc. 

Based on the evolutionary models, we have calculated the mass transfer rate predicted in the conservative case. Confirming the previous result (MD08), we found the theoretical mass transfer rate is $\lesssim 1.7\times 10^{-9}\msun$ yr$^{-1}$ for the donor masses $\leq 1.4\msun$. This transfer rate is much below the average mass accretion rate estimated from the long-term X-ray light curve.

We show that the discrepancy between the transfer rate and the accretion rate can be solved in two ways. One solution invokes strong irradiation of the donor by the X-rays from accretion. This may temporarily enhance the mass transfer rate. We have performed calculations of this effect and found that absorption of even a small fraction ($\sim$1 per cent) of the irradiating luminosity can cause an expansion of an outer layer of the donor and thus strongly enhance the mass transfer rate. We have estimated the time scale at which the mass transfer rate will vary and found it to be of $\sim$10 yr. Such variability in the transfer rate averaged over outbursts appears to be observed.

In the other solution, we have considered non-conservative mass transfer, in which a strong outflow carries away most of the mass and angular momentum flowing from the donor to the accretor. In our preferred solution, $\approx$70 per cent of the transfer mass escapes as an outflow and carries away the specific angular momentum of $\approx$1.2 of that of the donor. However, we note that no evidence for outflows has yet been found in GX 339--4. Thus, the irradiation solution appears preferable at this time. 

For both solutions, we calculated the predicted secular rate of the period change associated with the mass transfer, and obtained the expected secular change on a time scale of $P/\dot P\approx (3$--$5) \times 10^7$\,yr. If future measurements find the $\dot P$ to be substantially lower (or negative), this would imply the presence of a competing mechanism, e.g., magnetic braking.

\section*{Acknowledgements}

We thank Charles Bailyn, Jorge Casares, Phil Charles, Jean-Marie Hameury, Marianne Heida, Jean-Pierre Lasota, Alexey Pamyatnykh and Raghu Rao for valuable discussions. We also thank the referee for valuable suggestions. This research has been supported in part by the Polish National Science Centre grants 2013/10/M/ST9/00729 and 2015/18/A/ST9/00746.

\label{lastpage}

\begin{thebibliography}{}

\bibitem[\protect\citeauthoryear{Antoniadis et al.}{2012}]{antoniodasis12} Antoniadis J., van Kerkwijk M.~H., Koester D., Freire P.~C.~C., Wex N., Tauris T.~M., Kramer M., Bassa C.~G., 2012, MNRAS, 423, 3316 

\bibitem[\protect\citeauthoryear{Ball, Tout \& {\.Z}ytkow}{Ball et al.}{2012}]{ball12} Ball W.~H., Tout C.~A., {\.Z}ytkow A.~N., 2012, MNRAS, 421, 2713 

\bibitem[\protect\citeauthoryear{Basak \& Zdziarski}{2016}]{basak16}
Basak R., Zdziarski A. A., 2016, MNRAS, 458, 2199

\bibitem[\protect\citeauthoryear{Beech}{1988}]{beech88} Beech M., 1988, Ap\&SS, 147, 219 

\bibitem[\protect\citeauthoryear{Bernardini et al.}{2016}]{bernardini16} Bernardini F., Russell D.~M., Shaw A.~W., Lewis F., Charles P.~A., Koljonen K.~I.~I., Lasota J.~P., Casares J., 2016, ApJ, 818, L5 

\bibitem[\protect\citeauthoryear{Bhargava et al.}{2017}]{bhargava17} Bhargava Y., et al., 2017, ApJ, 849, 141 

\bibitem[\protect\citeauthoryear{Buxton et al.}{2012}]{buxton12} Buxton M.~M., Bailyn C.~D., Capelo H.~L., Chatterjee R., Din{\c c}er T., Kalemci E., Tomsick J.~A., 2012, AJ, 143, 130 

\bibitem[\protect\citeauthoryear{Coriat, Fender \& Dubus}{Coriat et al.}{2012}]{coriat12}
Coriat M., Fender R. P., Dubus G., 2012, MNRAS, 424, 1991

\bibitem[\protect\citeauthoryear{Cromartie et al.}{2019}]{cromartie19} Cromartie H. T., et al., 2019, arXiv:1904.06759 

\bibitem[\protect\citeauthoryear{Davis et al.}{2005}]{davis05} Davis S.~W., Blaes O.~M., Hubeny I., Turner N.~J., 2005, ApJ, 621, 372 

\bibitem[\protect\citeauthoryear{Dubus, Hameury \& Lasota}{Dubus et al.}{2001}]{dubus01} Dubus G., Hameury J.-M., Lasota J.-P., 2001, A\&A, 373, 251 

\bibitem[\protect\citeauthoryear{Eggleton}{1983}]{eggleton83} Eggleton P.~P., 1983, ApJ, 268, 368

\bibitem[\protect\citeauthoryear{Eggleton, Faulkner \& Cannon}{Eggleton et al.}{1998}]{eggleton98} Eggleton P.~P., Faulkner J., Cannon R.~C., 1998, MNRAS, 298, 831 

\bibitem[\protect\citeauthoryear{Flannery \& Ulrich}{1977}]{flannery77} Flannery B.~P., Ulrich R.~K., 1977, ApJ, 212, 533

\bibitem[\protect\citeauthoryear{Frank, King \& Lasota}{Frank et al.}{1992}]
{frank92} Frank J., King A. R., Lasota J. P., ApJ, 385, 45

\bibitem[\protect\citeauthoryear{Gies \& Bolton}{1986}]{gies86} Gies D.~R., Bolton C.~T., 1986, ApJ, 304, 371 

\bibitem[\protect\citeauthoryear{Gonz{\'a}lez Hern{\'a}ndez et al.}{2017}]{gonzalez17} Gonz{\'a}lez Hern{\'a}ndez J.~I., Su{\'a}rez-Andr{\'e}s L., Rebolo R., Casares J., 2017, MNRAS, 465, L15 

\bibitem[\protect\citeauthoryear{Hameury et al.}{1993}]{HKLR93}
Hameury J. M., King, A. R., Lasota J. P., Raison F., 1993, A\&A,
277, 81

\bibitem[\protect\citeauthoryear{Heida et al.}{2017}]{heida17}
Heida M., Jonker P. G., Torres M. A. P., Chiavassa A., 2017, ApJ,
846, 132 (H17)

\bibitem[\protect\citeauthoryear{Hynes et al.}{2003}]{hynes03}
Hynes R. I., Steeghs D., Casares J., Charles P. A., O'Brien, K.,
2003, ApJ, 583, L95

\bibitem[\protect\citeauthoryear{Islam \& Zdziarski}{2018}]{islam18} Islam N., Zdziarski A.~A., 2018, MNRAS, 481, 4513 

\bibitem[\protect\citeauthoryear{J{\c e}drzejec}{1969}]{jed69}
J{\c e}drzejec E., 1969, M.S.\ thesis, Warsaw Univ., unpublished

\bibitem[\protect\citeauthoryear{Jones et al.}{2011}]{jones11} Jones M.~I., Jenkins J.~S., Rojo P., Melo C.~H.~F., 2011, A\&A, 536, 71 

\bibitem[\protect\citeauthoryear{Justham, Rappaport \& Podsiadlowski}{Justham et al.}{2006}]{justham06} Justham S., Rappaport S., Podsiadlowski P., 2006, MNRAS, 366, 1415 

\bibitem[\protect\citeauthoryear{King}{1993}]{king93}
King A.R., 1993, MNRAS, 260, L5

\bibitem[\protect\citeauthoryear{King et al.}{1996}]{KFKR96}
King A. R., Frank J., Kolb U., Ritter H., 1996, ApJ, 467, 761

\bibitem[\protect\citeauthoryear{King et al.}{1997}]{KFKR97}
King A. R., Frank J., Kolb U., Ritter H., 1997, ApJ, 482, 919

\bibitem[\protect\citeauthoryear{Kolehmainen \& Done}{2010}]{kolehmainen10} Kolehmainen M., Done C., 2010, MNRAS, 406, 2206 

\bibitem[\protect\citeauthoryear{Koleva \& Vazdekis}{2012}]{koleva12} Koleva M., Vazdekis A., 2012, A\&A, 538, A143 

\bibitem[\protect\citeauthoryear{Koljonen \& Maccarone}{2017}]{koljonen17} Koljonen K.~I.~I., Maccarone T.~J., 2017, MNRAS, 472, 2181 

\bibitem[\protect\citeauthoryear{Lasota, Narayan \& Yi}{Lasota et al.}{1996}]{lasota96} 
Lasota J.-P., Narayan R., Yi I., 1996, A\&A, 314, 813 

\bibitem[\protect\citeauthoryear{Lattimer \& Prakash}{2010}]{lp10} 
Lattimer J.~M., Prakash M., 2010, in Lee S., ed., From Nuclei to Stars:
Festschrift in Honor of Gerald E Brown. World Scientific Press, Singapore, p. 275 (arXiv:1012.3208). 

\bibitem[\protect\citeauthoryear{Linares, Shahbaz \& Casares}{Linares et al.}{2018}]{linares18} Linares M., Shahbaz T., Casares J., 2018, ApJ, 859, 54 

\bibitem[\protect\citeauthoryear{Luck}{2017}]{luck17} Luck R.~E., 2017, AJ, 153, 21 

\bibitem[\protect\citeauthoryear{Markert et al.}{1973}]{markert73}
Markert T. H., Canizares C. R., Clark G. W., Lewin W. H. G.,
Schnopper H. W., Sprott G. F., 1973, ApJL, 184, L67

\bibitem[\protect\citeauthoryear{McDonald, Zijlstra \& Watson}{McDonald et al.}{2017}]{mcdonald17} McDonald I., Zijlstra A.~A., Watson R.~A., 2017, MNRAS, 471, 770 

\bibitem[\protect\citeauthoryear{Mendez \& van der Klis}{1997}]{mendez97}
Mendez M., van der Klis M., 1997, 1997, ApJ, 479, 926

\bibitem[\protect\citeauthoryear{Mu{\~n}oz-Darias, Casares \& Mart{\'{\i}}nez-Pais}{Mu{\~n}oz-Darias et al.}{2008}]{munoz08}
Mu{\~n}oz-Darias T., Casares J., Mart{\'{\i}}nez-Pais I. G., 2008, MNRAS,
385, 2205 (MD08)

\bibitem[\protect\citeauthoryear{Mu{\~n}oz-Darias et al.}{2013}]{munoz13} Mu{\~n}oz-Darias T., Coriat M., Plant D.~S., Ponti G., Fender R.~P., Dunn R.~J.~H., 2013, MNRAS, 432, 1330 

\bibitem[\protect\citeauthoryear{{\"O}zel et al.}{2010}]{ozel10} {\"O}zel F., Psaltis D., Narayan R., McClintock J.~E., 2010, ApJ, 725, 1918 

\bibitem[\protect\citeauthoryear{{\"O}zel \& Freire}{2016}]{ozel16} {\"O}zel F., Freire P., 2016, ARA\&A, 54, 401 

\bibitem[\protect\citeauthoryear{Paczy{\'n}ski}{1967}]{paczynski67}
Paczy{\'n}ski B., 1967, Acta Astron., 17, 287

\bibitem[\protect\citeauthoryear{Paczy{\'n}ski}{1969}]{paczynski69} Paczy{\'n}ski B., 1969, Acta Astron., 19, 1 

\bibitem[\protect\citeauthoryear{Paczy{\'n}ski}{1970}]{paczynski70} Paczy{\'n}ski B., 1970, Acta Astron., 20, 47 

\bibitem[\protect\citeauthoryear{Paczy\'nski \& Sienkiewicz}{1972}]{ps72}
Paczy\'nski B., Sienkiewicz R., 1972, Acta Astron., 22, 73

\bibitem[\protect\citeauthoryear{Pamyatnykh et al.}{1998}]{pamyatnykh98} Pamyatnykh A.~A., Dziembowski W.~A., Handler G., Pikall H., 1998, A\&A, 333, 141 

\bibitem[\protect\citeauthoryear{Parker et al.}{2016}]{parker16}
Parker M. L., et al., 2016, ApJ, 821, L6

\bibitem[\protect\citeauthoryear{Podsiadlowski}{1991}]{podsiadlowski91}
Podsiadlowski Ph., 1991, Nature 350, 136

\bibitem[\protect\citeauthoryear{Ponti et al.}{2012}]{ponti12} Ponti G., Fender R.~P., Begelman M.~C., Dunn R.~J.~H., Neilsen J., Coriat M., 2012, MNRAS, 422, L11

\bibitem[\protect\citeauthoryear{Poutanen \& Veledina}{2014}]{pv14} Poutanen J., Veledina A., 2014, SSRv, 183, 61 

\bibitem[\protect\citeauthoryear{Prugniel, Vauglin \& Koleva}{Prugniel et al.}{2011}]{prugniel11} Prugniel P., Vauglin I., Koleva M., 2011, A\&A, 531, 165 

\bibitem[\protect\citeauthoryear{Rappaport, Joss \& Webbink}{Rappaport et al.}{1982}]{rappaport82} Rappaport S., Joss P.~C., Webbink R.~F., 1982, ApJ, 254, 616

\bibitem[\protect\citeauthoryear{Reffert et al.}{2015}]{reffert15} Reffert S., Bergmann C., Quirrenbach A., Trifonov T., K{\"u}nstler A., 2015, A\&A, 574, 116 

\bibitem[\protect\citeauthoryear{Ritter}{2008}]{ritter08}
Ritter H., 2008, NewAR, 51, 869

\bibitem[\protect\citeauthoryear{Ritter, Zhang \& Kolb}{Ritter et al.}{2000}]{RZK00}
Ritter H., Zhang Z.-Y., Kolb U., 2000, A\&A, 360, 969

\bibitem[\protect\citeauthoryear{Savonije}{1983}]{Savonije83}
Savonije G. J., 1983, in Lewin W. H. G., van den Heuvel E. P. J.,
eds, Accretion-driven Stellar X-ray Sources. Cambridge Univ. Press,
Cambridge, p. 343

\bibitem[\protect\citeauthoryear{Sch{\"o}nberg \& Chandrasekhar}{1942}]{sc42}
Sch{\"o}nberg M., Chandrasekhar S., 1942, ApJ, 96, 161

\bibitem[\protect\citeauthoryear{Shahbaz, Fender \& Charles}{Shahbaz et al.}{2001}]{shahbaz01}
Shahbaz T., Fender R., Charles P. A., 2001, A\&A 376, L17

\bibitem[{{Shakura} \& {Sunyaev}(1973)}]{ss73}
{Shakura} N.~I., {Sunyaev} R.~A., 1973, \aap, 24, 337

\bibitem[\protect\citeauthoryear{Shaw et al.}{2019}]{shaw19} Shaw A.~W., Tetarenko B.~E., Dubus G., Din{\c c}er T., Tomsick J.~A., Gandhi P., Plotkin R.~M., Russell D.~M., 2019, MNRAS, 482, 1840

\bibitem[\protect\citeauthoryear{van Kerkwijk, Breton \& Kulkarni}{van Kerkwijk et al.}{2011}]{vankerkwijk11} van Kerkwijk M.~H., Breton R.~P., Kulkarni S.~R., 2011, ApJ, 728, 95 

\bibitem[\protect\citeauthoryear{Verbunt}{1993}]{verbunt93} Verbunt F., 1993, ARA\&A, 31, 93

\bibitem[\protect\citeauthoryear{Vilhu, Ergma \& Fedorova}{Vilhu et al.}{1994}]{VEF94}
Vilhu O., Ergma E., Fedorova A., 1994, A\&A, 291, 842

\bibitem[\protect\citeauthoryear{Webbink, Rappaport \& Savonije}{Webbink et al.}{1983}]{webbink83}
Webbink R. F., Rappaport, S. A., Savonije, G. J., 1983, ApJ, 270,
678

\bibitem[\protect\citeauthoryear{Zdziarski et al.}{1998}]{zdziarski98} Zdziarski A.~A., Poutanen J., Miko{\l}ajewska J., Gierli{\'n}ski M., Ebisawa K., Johnson W.~N., 1998, MNRAS, 301, 435 

\bibitem[\protect\citeauthoryear{Zdziarski et al.}{2004}]{zdziarski04}
Zdziarski A. A., Gierli\'nski M., Miko{\l}ajewska J., Wardzi\'nski
G., Smith D. M., Harmon B. A., Kitamoto S., 2004, MNRAS, 351, 791 (Z04)

\bibitem[\protect\citeauthoryear{Zdziarski et al.}{2007}]{ZWG07}
Zdziarski A. A., Wen, L., Gierli\'nski M., 2007, MNRAS, 377, 1006

\bibitem[\protect\citeauthoryear{Zdziarski, Miko{\l}ajewska \& Belczy{\'n}ski}{Zdziarski et al.}{2013}]{zmb13}
Zdziarski A.~A., Miko{\l}ajewska J., Belczy{\'n}ski K., 2013, MNRAS, 429, L104  

\bibitem[\protect\citeauthoryear{Zdziarski et al.}{2016}]{z16}
Zdziarski A.~A., Zi{\'o}{\l}kowski J., Bozzo E., Pjanka P., 2016,
A\&A, 595, 52

\bibitem[\protect\citeauthoryear{Zi\'o{\l}kowski}{2005}]{ziolkowski05}
 Zi\'o{\l}kowski J., 2005, MNRAS, 358, 851

\bibitem[\protect\citeauthoryear{Zi\'o{\l}kowski \& Zdziarski}{2017}]{zz17}
 Zi\'o{\l}kowski J., Zdziarski A. A., 2017, MNRAS, 469, 3315

 \bibitem[\protect\citeauthoryear{Zi\'o{\l}kowski \& Zdziarski}{2018}]{zz18}
 Zi\'o{\l}kowski J., Zdziarski A. A., 2018, MNRAS, 480, 1580

\end{thebibliography}
\end{document}